\documentclass[conference]{IEEEtran}

\usepackage{helvet}
\usepackage{courier}
\usepackage{graphicx}
\usepackage{times}
\usepackage{latexsym}
\usepackage{amsmath}
\usepackage{multirow}
\usepackage{url}
\usepackage{paralist}
\usepackage{color} 
\usepackage[normalem]{ulem} 

\ifCLASSINFOpdf
\else
\fi
\begin{document}
%
\title{Exploring Image Virality in Google Plus}

\author{\IEEEauthorblockN{Marco Guerini}
\IEEEauthorblockA{Trento RISE\\
Trento - Italy\\
Email: marco.guerini@trentorise.eu}
\and
\IEEEauthorblockN{Jacopo Staiano}
\IEEEauthorblockA{University of Trento\\
Trento - Italy\\
Email: staiano@disi.unitn.it}
\and
\IEEEauthorblockN{Davide Albanese}
\IEEEauthorblockA{Fondazione Edmund Mach, CRI-CBC\\
San Michele all’Adige (TN) - Italy\\
Email: davide.albanese@fmach.it}}


%


\maketitle

\begin{abstract}
Reactions to posts in an online social network show different dynamics depending on several textual features 
of the corresponding content. Do similar dynamics exist when images are posted? 
Exploiting a novel dataset of posts, gathered from the most popular 
Google+ users, we try to give an answer to such a question. 
We describe several virality phenomena that emerge when taking into account visual characteristics of images 
(such as orientation, mean saturation, etc.).  
We also provide hypotheses and potential explanations for the dynamics behind
them, and include cases for which common-sense expectations do not 
hold true in our experiments.
\end{abstract}


%
\IEEEpeerreviewmaketitle

\section{Introduction}

How do things become `viral' on the Internet? And what exactly do we mean by `influence'? 
Since marketing and industry  people want their messages to spread in the most effective 
and efficient way possible, these questions have received a great deal of attention, particularly in recent years, 
as we have seen a dramatic growth of  social networking on the Web. Generally speaking, virality refers to the 
tendency of a content either to spread quickly within a community or to receive a great deal of attention by it. 
In studying the spreading process we will focus on the content and its characteristics, rather than on the structure 
of the network through which the information is moving. In particular, we will investigate the relationships between visual characteristics 
-- of images enclosed in Google+ posts -- and virality
phenomena. We will use three virality metrics: plusoners, replies and resharers.

This exploratory work stems from the 
use people make of social networking websites such as Google+, Facebook 
and similar:  
we hypothesized that perceptual characteristics of an image could indeed affect the virality of the post 
embedding it, and that -- for example -- cartoons, panorama or self-portraits picture affect users' reactions in different ways. 
The aim of this paper is to investigate whether signs of such ``common-sense'' intuition emerge 
from large-scale data made available on popular social networking websites like Google+ and, in such
case, to open discussion on the associated phenomena.

The paper is structured as follows: first, we review previous works addressing the topic of virality in social networks, 
and particularly some focusing on content impact. 
Then, after describing the dataset collected and used for this work, 
we proceed with the study of virality of Google+ posts and the characteristics of their content.
We  also discuss the behavior of virality indexes in terms of their alternative use, arguing that 
plusones and comments fulfill a similar purpose of followers' ``appreciation" while reshares have a different role of ``self-representation".  
Finally, we investigate possible interactions between image characteristics  and users' typology, 
in order to understand to what extent results are generalizable or typical of a community, gathered around a common interest. 

\section{Related Works}

Several researchers studied information flow, community building and similar processes using Social Networking sites 
as a reference \cite{contagionDigg,digging,caseStudyOnComments,credibility}. However, the great 
majority concentrates on network-related features without taking into account the actual content spreading within the network \cite{voting}.
A hybrid approach focusing on both product characteristics and network related features is presented in \cite{aral2011creating}: 
the authors study the effect of passive-broadcast and active-personalized notifications embedded in an application 
to foster word of mouth.

Recently, the correlation between content characteristics and virality has begun to be investigated, especially 
with regard to textual content; in \cite{opinion}, for example,
features derived from sentiment analysis of comments are used to predict the popularity of stories. 
The work presented in \cite{virality} uses \textit{New York Times}' articles to examine the relationship between 
emotions evoked by the content and virality, using semi-automated sentiment analysis to quantify the affectivity 
and emotionality of each article. Results suggest a strong relationship between affect and virality; still, the 
virality metric considered is interesting but very limited: it only consists of how many people emailed the article.  The relevant work in \cite{danescu2012you} measures a different form of content 
spreading by analyzing which are the features of a movie quote that make it ``memorable" online. 
Another approach to content virality, somehow complementary to the previous one, is presented in \cite{simmons2011memes}, 
trying to understand which modification dynamics make a meme spread from one person to another 
(while movie quotes spread remaining exactly the same).  

More recently, some works tried to investigate how different textual contents give rise to different reactions 
in the audience: the work presented in \cite{marco:carlo:gozde:ICWSM-11}  correlates several viral 
phenomena with the wording of a post,  while \cite{guerini2012linguistic} show that  specific content features 
variations (like the readability level of an abstract) differentiate among virality level of downloads, 
bookmarking, and citations. Still, to our knowledge, no attempt has been made yet to investigate the relation between visual content 
characteristics and virality.

\section{Data Description}

Using the Google+ API\footnote{https://developers.google.com/+/api/}, we harvested the public posts from the 
979 top followed users in Google+
(\texttt{plus.google.com}), as reported by the \texttt{socialstatistics.com} website on March 2nd 2012\footnote{The dataset presented and used in this work will be made available
to the community for research purposes.}.
The time span for the harvesting is one year, from June 28th 2011 (Google+ date of launch) to June 29th 2012.

We decided to focus on the most popular users for several reasons: (i) the dataset is uniform from the point of view of sample role, i.e. VIPs, 
(ii) the behavior of the followers is consistent -- e.g. no friendship dynamics -- and (iii) extraneous effects due to followers network is minimized, since 
top followed users' network is vast enough to grant that, if  a content is viral, a certain amount of reactions will be obtained.

We defined 3 subsets of our dataset, comprising respectively: (i) posts containing a static image, (ii) posts containing an animated image (usually, \texttt{gif}), 
(iii) posts without attachments (text-only). All other posts (containing as attachment videos, photo albums, links to external sources) were discarded. Statistics for our dataset are reported in Table \ref{dataset-stats}. 
For each post, we considered three virality metrics\footnote{Since the API provide only an aggregate number, we cannot make any temporal analysis of how reactions to a post were accumulated over time.}: 

\begin{itemize}
 \item \textbf{Plusoners}: the number of people who +1'd;
 \item \textbf{Replies}: the number of comments;
 \item \textbf{Resharers}: the number of people who reshared.
\end{itemize}

\begin{table}[htdp]
\caption{An overview of the Google+ dataset.}
\begin{center}
\begin{tabular}{ll}
\hline
\multicolumn{2}{c}{\textbf{Global}}\\
\hline
actors & 979 \\
posts & 289434  \\
published interval & 6/28/11--6/29/12\\
\hline
\multicolumn{2}{c}{\textbf{Posts with static images}}\\
\hline
actors & 950 \\
posts & 173860 \\
min/max/median posts per actor & 1/3685/ 65.5\\
min/max/median plusoners per post & 0/9703/33.0 \\
min/max/median replies per post$^a$ & 0/571/12.0 \\
min/max/median resharers per post & 0/6564/4.0 \\
\hline
\multicolumn{2}{c}{\textbf{Posts with animated images}}\\
\hline
actors & 344 \\
posts: & 12577 \\
min/max/median posts per actor & 1/2262/3.0 \\
min/max/median plusoners per post & 0/5145/17.0 \\
min/max/median replies per post & 0/500/7.0 \\
min/max/median resharers per post & 0/6778/10.0 \\
\hline
\multicolumn{2}{c}{\textbf{Posts without attachments, text-only}}\\
\hline
actors & 939 \\
posts & 102997 \\
min/max/median posts per actor & 1/1744/41.0 \\
min/max/median plusoners per post & 0/20299 /16.0 \\
min/max/median replies per post & 0/538/17.0 \\
min/max/median resharers per post & 0/13566/1.0 \\
\hline
  \multicolumn{2}{c}{$^a${\scriptsize Replies count is cut around 500 by the API service.}}\\
\end{tabular}
\end{center}
\label{dataset-stats}
\end{table}%

\begin{figure}[h!] 
\begin{center}
\includegraphics[width=1\columnwidth]{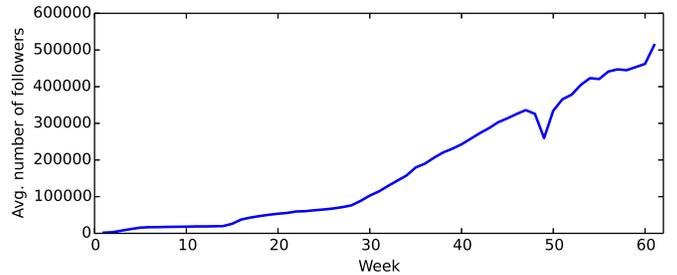}
\caption{Average number of followers per user, at 1-week temporal granularity.}
\label{fig:timelinefriends} 
\end{center} 
\end{figure}

\begin{figure}[h!] 
\begin{center}
\includegraphics[width=1\columnwidth]{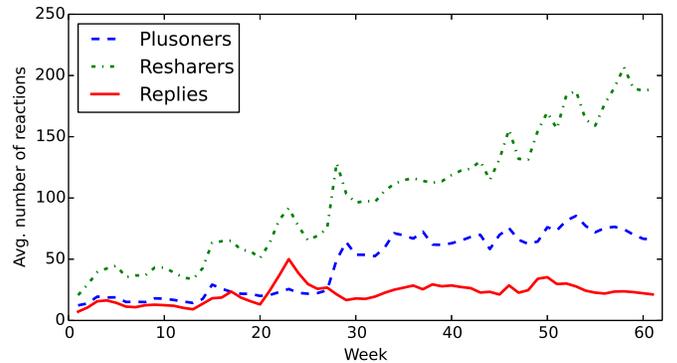}
\caption{Average number of reactions per user, at 1-week temporal granularity. This value represents the average number of reactions elicited by each
user's posts over 1-week time-slices.}
\label{fig:timelinereactions} 
\end{center} 
\end{figure}

In Figures~\ref{fig:timelinefriends} and \ref{fig:timelinereactions} we display the evolution over time of the network underlying our
dataset (using a week as temporal unit), and of the reactions to posts given by users, respectively.
We notice that:

\begin{enumerate}
 \item the average number of reactions per user shows quite different trends depending on the metric considered: 
 while replies tend not to be affected by the growth of the
network
, reshares and, to a lesser degree plusones, show an ever-growing trend.
 \item The temporal plot of the average number of followers per user (Figure~\ref{fig:timelinefriends}) in our dataset (in Google+ terminology, the number of people who
\emph{circled} them) shows a gradient increase around weeks
28/29. Interestingly, this is reflected in the plot of reactions over time (Figure~\ref{fig:timelinereactions}): the gradient increases around the same weeks, for reshares and
plusones; these effects are most probably due to Google+ transitioning from beta to public in late September 2011 (a similar phenomenon is reported also in ~\cite{Schioberg2012}).
 \item Finally, the orders of magnitude of such growths are very different: we notice that while reactions increase of a factor of ~7 over the time period we took into account, the
total number of followers increased of a factor of ~25.
\end{enumerate}


The relative amount of followers' reactions does not significantly increase as the network
grows\footnote{It has been noted how (see, for instance, http://on.wsj.com/zjRr06), especially in the time frame we
consider, users' activity did not increase much in front of the exploding network size.}.
As detailed in the next section, our analyses are based on comparing probability distributions: e.g. we evaluate if grayscale images have a significantly higher or smaller
probability of reaching a certain virality score than colored ones.
In the following analyses, for the sake of clarity, our discussion will not take into account the normalization factor (i.e. the size of the audience when a content is posted).
Indeed, we have run the same analyses normalizing the virality indexes of a given post against its \emph{potential} audience: i) the effects are still visible, ii) the
effects are consistent both in significance and sign with the not-normalized distributions, but iii) differences have lower magnitude (explained by the fact that virality indexes
should be normalized using the \emph{actual} audience -- e.g. the followers exposed to the content). 
Thus, since
we are interested in comparing the virality of different image categories and our preliminary experiments showed that by
normalizing the indexes their comparisons, their sign, and the derived interpretations still hold, we choose to report the non-normalized version of the results that are more
intuitively readable.

\begin{figure}[h!] 
\begin{center}
\includegraphics[width=1\columnwidth]{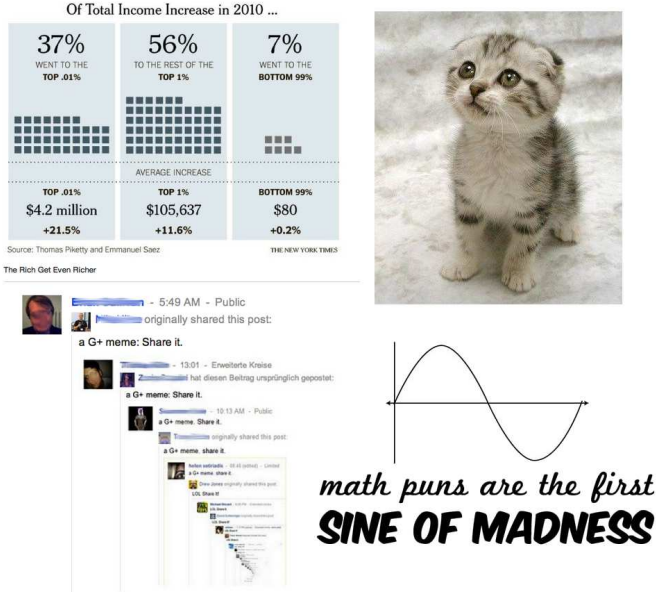}
\caption{Exemplar pictures from the dataset.}
\label{img:pics} 
\end{center} 
\end{figure}

In the following sections, after the analyses 
of text-only posts and of posts containing an animated image, we will consider the subset 
of static images as the reference dataset. Exemplar pictures taken from the dataset are shown in Figure \ref{img:pics}, 
depicting some image categories that we will take into account in the following sections.  

\section{Data Analysis}

Virality metrics in our dataset follow a power-law-like distribution thickening toward low virality score. In order to evaluate the 
``virality power" of the features taken into account, we compare the virality indexes in terms of empirical Complementary Cumulative Distribution Functions (CCDFs). 
These functions are commonly used to analyse online social networks in terms of growth in size and activity (see for example \cite{ahn2007,jiang2010}, or 
the discussion presented in \cite{leskovec2008dynamics}) and also for measuring content diffusion, e.g. the number of retweets
of a given content \cite{kwak2010}. Basically, these functions account for the probability $p$ that a virality index will be greater than $n$ and are defined as follows:

\begin{equation}
\hat{F}(n) = \frac{\text{number of posts with virality index} > n}{\text{total number of posts}} 
\end{equation}

For example, the
probability of having a post with more than 75 plusoners is indicated with $\hat{F}_{plus}(75) = \mathrm{P}
(\mathrm{\#plusoners}>75)$. 
In the following sections we use CCDFs to understand the relation between image characteristics 
and post virality; in order to assess whether the CCDFs of the several types of posts we take into account show significant differences, we will use the Kolmogorov-Smirnov (K--S)
goodness-of-fit test, which specifically targets cumulative distribution functions. 


\subsection{Image vs. text-only}

First of all, we aim to understand what is the impact of ``adding an image to a post" in Google+. Some studies \cite{tswhitepaper} 
already show that posts containing an
image are much more viral than simple plain-text posts, and that various characteristics of image based banners 
affect viewer's recall and clicks \cite{li1999cognitive}.  This finding can be explained in light of a ``rapid 
cognition" model \cite{ambady1992thin,kenny1994interpersonal}. In this
model, the user has to decide in a limited amount of time, and within a vast information flow of posts, whether 
to take an action on a particular post (e.g. to reply, reshare, give it a plusone). 
Thus, pictures, and the characteristics thereof analyzed in the following sections, might play a role of paramount importance in her decision-making process 
as she exploits visual cues that grab her attention.
In some respects, the rapid cognition model is
reminiscent of the mechanisms by which humans routinely make judgments about strangers' personality and 
behavior from very short 
behavioral sequences and non-verbal cues \cite{Lepri:2010:ESG:1891903.1891913,curhan2007thin}. 

\begin{figure}[h!] 
\begin{center}
\includegraphics[width=0.75\columnwidth]{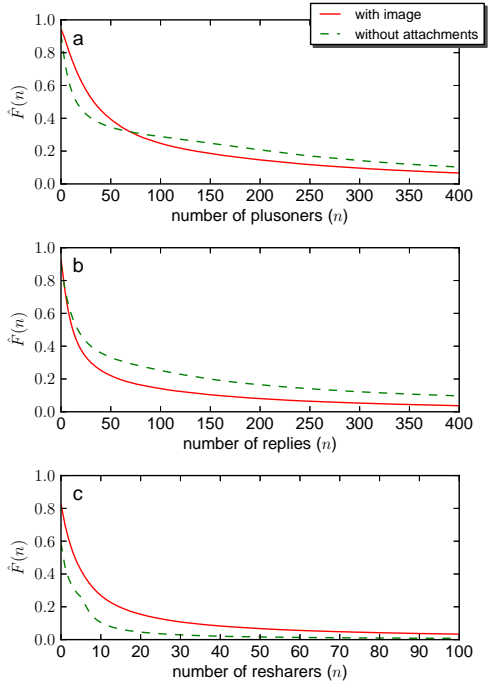}
\caption{Virality CCDFs for posts with image vs. text-only posts.}
\label{img:cdf_image_vs_noattach} 
\end{center} 
\end{figure}

In order to investigate the general impact of images we compared posts containing a picture with posts containing only text. 
While our findings overall coincide
with \cite{tswhitepaper}, some interesting phenomena emerged. First, we see that the probability for a 
post with an image to have a high number of resharers
is almost three times greater ($\hat{F}_{resh}(10)$ = 0.28 vs. 0.10, K--S test $p<0.001$), 
see Figure \ref{img:cdf_image_vs_noattach}.c. Still,  the CCDFs for the other virality indexes show different trends: 

\begin{itemize}
\item Posts containing images have lower probability of being viral when it comes to number of comments 
($\hat{F}_{repl}(50)$ = 0.33 vs. 0.22, K--S test $p<0.001$), see Figure \ref{img:cdf_image_vs_noattach}.b. This can be explained by the
fact that text-only posts elicit more ``linguistic-elaboration'' than images (we also expect that the average 
length of comments is higher for text-only posts but we do not
investigate this issue here).
\item Also, if we focus on simple appreciation (plusoners in Figure \ref{img:cdf_image_vs_noattach}.a), 
results are very intriguing: while up to about 75 plusoners the probability of having posts containing images is higher, after this threshold the 
situation capsizes. This finding can be of support to the hypothesis that, while 
it is easier to impress with images in the information flow --- as
argued with the aforementioned ``rapid cognition" model --- high quality textual content can impress more.  
\end{itemize}

\subsection{Static vs. Animated}
\label{sec:animated}

Animated images add a further dimension to pictures expressivity. Having been around since the beginning 
of the Internet (the \texttt{gif} format was introduced in late 80's), animated images have
had alternate fortune, especially after the wide spread of services like youtube and the availability of 
broadband. Nonetheless, they are still extensively used to produce simple
animations and short clips. Noticeably, the value of simple and short animations has been 
acknowledged by Twitter with the recently released \emph{Vine} 
service.  

\begin{figure}[h!] 
\begin{center}
\includegraphics[width=0.75\columnwidth]{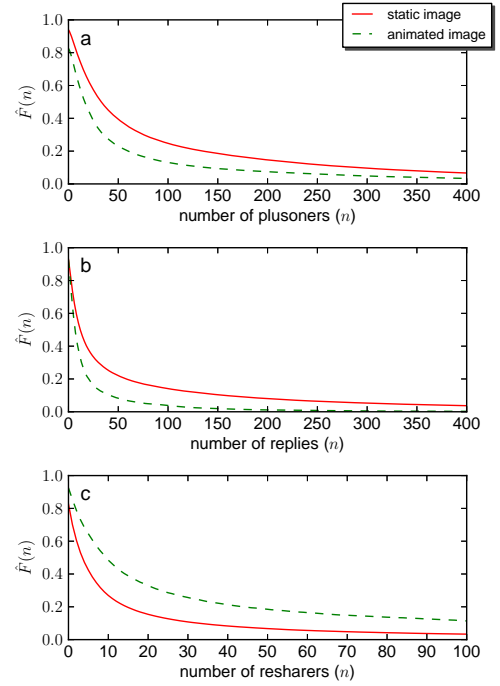}
\caption{Virality CCDFs for static vs. animated images.}
\label{img:cdf_static_animated} 
\end{center} 
\end{figure}

Whether a post contains a static or animated image has a strong discriminative impact on all virality indexes, see Figure \ref{img:cdf_static_animated}. With respect to plusoners
and replies,
static images tend to show higher CCDFs (respectively two and three times more, $\hat{F}_{plus}(75)$ = 0.30 vs. 0.17, $\hat{F}_{repl}(50)$ = 0.22 vs. 0.08,  K--S test $p<0.001$), 
while on resharers the opposite holds. 

The fact that $\hat{F}_{resh}(n)$ is two times higher for posts containing animated images  ($\hat{F}_{resh}(10)$ = 0.48 vs. 0.27, K--S test $p<.001$) can be potentially explained
by the fact that animated images are usually built to convey a small 
 ``memetic'' clip - i.e. \emph{funny}, \emph{cute} or \emph{quirky} situations as suggested in \cite{dufour2011investigation}. 
 
In order to verify this hypothesis we have annotated
a small 
random subsample of 200 images. 81\% of these animated images were found to be ``memetic'' (two 
annotators were used, positive example if the image score 1 at least on one of the aforementioned  dimensions, annotator agreement is very high --- Cohen's kappa 0.78). These
findings indicate that animated images are mainly a vehicle for amusement, at least on Google+. 

\subsection{Image Orientation}

We then focused 
on the question whether image orientation (\emph{landscape}, \emph
{portrait} and \emph{squared}) has any impact on virality
indexes. 
We included squared images in our analysis since they are typical of popular services  a la \emph{Instagram}. 
These services enable users to apply digital filters to the pictures they take and
confine photos to a squared shape, similar to Kodak Instamatic and Polaroids, providing a so-called
``vintage effect".

\begin{figure}[h!] 
\begin{center}
\includegraphics[width=0.75\columnwidth]{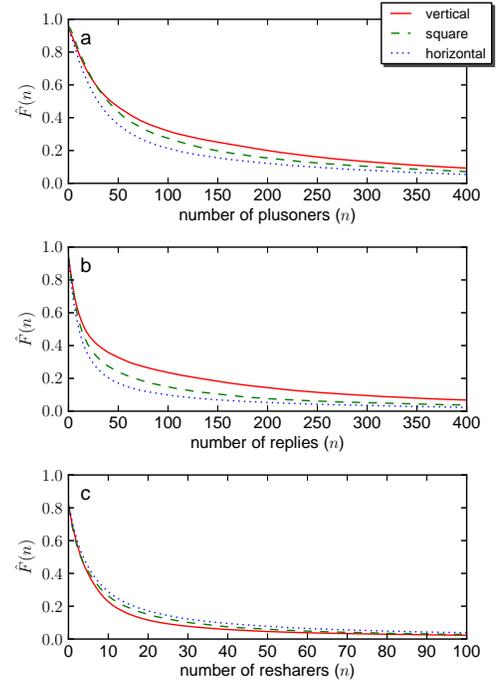}
\caption{Virality CCDFs for image orientation.}
\label{img:cdf_image_orientation} 
\end{center} 
\end{figure} 

We have annotated a small random subsample of 200 images. 55\% of these images were found to 
be ``Instagrammed'' (two annotators were used,
positive example if the image is clearly recognized as modified with a
filter; annotator agreement is high -- Cohen's kappa 0.68). Note that,
if we include also black and white squared pictures without any other
particular filter applied (b/w is one of the "basic" filter provided
by \emph{Instagram}) the amount of Instagrammed pictures rises to 65\%.
Obviously, the ratio of pictures modified with this and
similar services could be higher; here, we rather wanted to
identify those pictures that were clearly recognized as seeking for
the aforementioned ``vintage effect". 

While the orientation seems not to have strong impact on resharers, 
with a mild prevalence of horizontal pictures (see Figure \ref{img:cdf_image_orientation}.c), 
plusoners and replies tend to well discriminate among
various image orientations. In particular, portrait images show higher probability of being viral than 
squared images than, in turn, landscapes (see Figure
\ref{img:cdf_image_orientation}.a and \ref{img:cdf_image_orientation}.b). 

Furthermore, CCDFs indicate that vertical images
tend to be more viral than horizontal ones ($\hat{F}_{plus}(75)$ = 0.38 vs. 0.26, $\hat{F}_{repl}(50)$ = 0.38 vs. 0.17,  K--S test $p<0.001$).
Hence, while squared images place themselves in the middle in any metric, landscape images 
have lower viral probability for plusones and replies but slightly higher probability for reshares. 

This can be partially explained by the fact that we are analyzing ``celebrities'' posts. If the vertically-orientated image contains the portrait of a celebrity this is
more likely to be appreciated rather that reshared, since the act of resharing can also be seen as a form of ``self-representation'' of the follower 
(we will analyze the impact of pictures containing faces in the following section).
The opposite holds for landscapes, i.e. they are more likely to be reshared and used for self-representation.

%
\subsection{Images containing one face} 
\label{sec:faces}

In traditional mono-directional media (e.g. tv, billboards, etc.) a widely used promotion strategy is the use of 
testimonials, especially celebrities endorsing a product. Is the
same strategy applicable to Social Media? Understanding the effect of posting images with faces by most 
popular Google+ users  (and hypothesizing  that those are their faces) is a first step in the direction of finding an answer.  

We computed how many faces are found in the images, along with the ratio of the area that
include faces and the whole image area, using the Viola-Jones~\cite{Viola2004} face detection algorithm. 
We considered images containing one face vs. images containing no faces. We did not consider the surface of 
image occupied by the face (i.e. if it is a close-up portrait, or just a
small face within a bigger picture). The discriminative effect of containing a face on virality is statistically significant but small.
Still, the pictures containing faces tend to have mild effect on resharers 
(slightly higher replies and plusoners but lower resharers as compared to images with no faces).

\begin{figure}[h!] 
\begin{center}
\includegraphics[width=0.75\columnwidth]{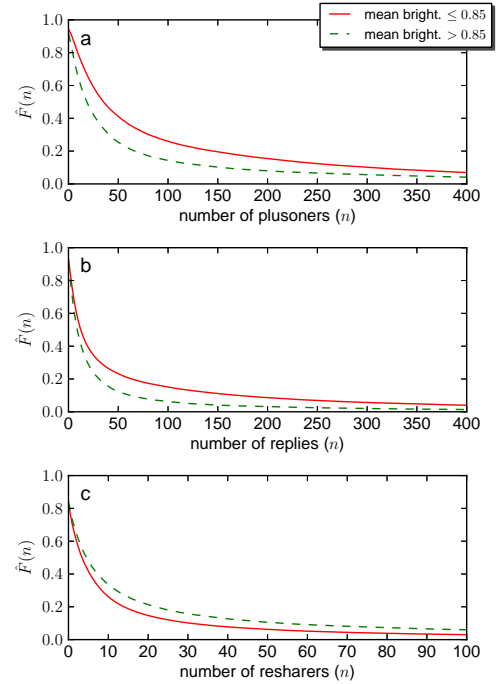}
\caption{Virality CCDFs for image Brightness.}
\label{img:cdf_brightness} 
\end{center} 
\end{figure} 

In order to verify the hypothesis mentioned earlier, i.e. that self-portraits tend to be reshared less, we also focused on a subsample 
of images containing faces that cover at least 10\% of the image surface (about 6400 instances). In this case, the differences among 
indexes polarize a little more (higher plusoners and comments, lower resharers), as we were expecting. Unfortunately, images 
with even higher face/surface ratio are too few to further verify the hypotheses. 

\subsection{Grayscale vs. Colored} 
\label{Grayscale} 

The impact and meaning of black-and-white (i.e. grayscale) photographic images has been studied from different perspectives (e.g. semiotics and psychology)
and with reference to different fields (from documentary to arts and advertising).
Rudolf Arnheim, for example, argues that color produces essentially \emph{emotional} experience, whereas shape corresponds 
to \emph{intellectual} pleasure  \cite{arnheim1987art}. 
Hence, black-and-white photography, because of its absence of
expressive colors, focuses on shapes that require intellectual
reflection and brings to explore aesthetic possibilities. 
We want to understand if such functions and effects can be spotted in our virality indexes. 

In order to have a ``perceptual" grayscale (some images may contain highly desaturated colors and so perceived as shades of gray) 
we dichotomized the dataset according to the mean-saturation index of the images, using a very conservative threshold of 0.05 (on a 0-1 scale).  

\begin{figure}[h!] 
\begin{center}
\includegraphics[width=0.75\columnwidth]{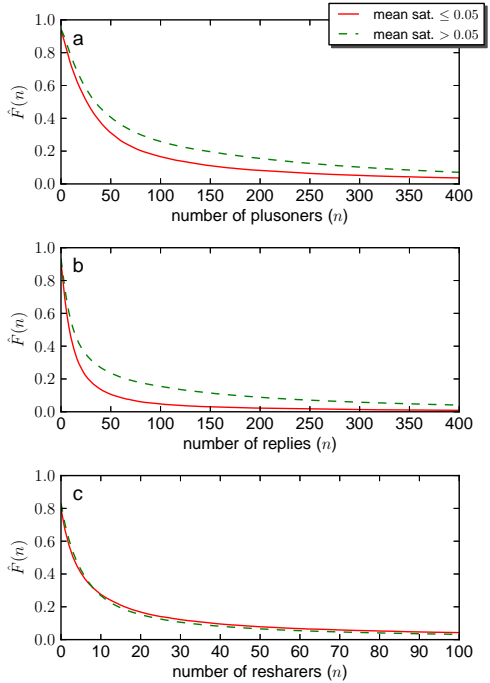}
\caption{Virality CCDFs for Grayscale vs. Colored images.}
\label{img:cdf_saturation} 
\end{center} 
\end{figure} 

As can be seen in Figure \ref{img:cdf_saturation}.a 
and \ref{img:cdf_saturation}.b, colored images (with saturation higher than 0.05) 
have a higher probability of  collecting more 
plusoners and replies as compared to images with lower saturation (grayscale). In particular the probability functions 
for replies is more than two times higher ($\hat{F}_{repl}(50)$ values are 0.26 vs. 0.10, K--S test $p<0.001$). 
Instead, image saturation has no relevant impact on resharers.

\subsection{Very Bright Images} 
\label{Bright-Images} 
After converting each image in our dataset to the HSB color space, we extracted its mean Saturation and Brightness.
More in detail, the HSB (Hue/Saturation/Brightness) color space describes each pixel in an image as a point on 
a cylinder: the Hue dimension representing its color within the set
of primary-secondary ones, while Saturation and Brightness describe respectively how close to the pure color 
(i.e. its Hue), and how bright it is.We split the dataset according to images mean brightness using a threshold of 0.85 (in a scale included between 0 and 1).
Usually images with such an high mean brightness tend to be cartoon-like images rather than pictures.  Previous research ~\cite{Ianeva2003} 
has shown that pixel brightness is expected to be higher in cartoon-like (or significantly ``photoshopped') than in natural images.


Image brightness level 
has a strong impact on plusoners and replies, and a milder one on resharers. Brighter
images have a lower probability of being viral on the first two indexes (Figure \ref{img:cdf_brightness}.a and \ref{img:cdf_brightness}.b) and a higher probability on the latter
(Figure \ref{img:cdf_brightness}.c).  
In particular, lower brightness images have a plusone and reshare probability almost two times higher 
($\hat{F}_{plus}(75)$ = 0.31 vs. 0.18, $\hat{F}_{repl}(50)$ = 0.23 vs. 0.12,  K--S test $p<0.001$), while for resharers it is 27\% 
higher in favor of high brightness images ($\hat{F}_{resh}(10)$ = 0.33 vs. 0.26).   
Surprisingly, analyzing a small random subsample of 200 very bright images, we found that  while 88\% of these images contained some text, 
as we would have expected, only 13\% were cartoon/comics and only 13\% contained the real picture of an object as subject, even if 
highly "photoshopped". Above all, only a small amount of these images (21\%) was considered funny or memetic\footnote{Two 
annotators were used, four binary categories were provided (contain-text/comics/real-picture-obj/funny). The overall inter-annotator agreement on these categories is high, Cohen's
kappa 0.74.}.
The great majority comprised pictures containing infographics, screenshots of software programs, screenshots of social-networks posts and similar. 
In this respect we are analyzing a content that is meant to be mainly informative, and is somehow complementary to  the content 
of animated pictures (mainly intended for amusement, see~\ref{sec:animated}).

%

\subsection{Vertical and Horizontal edges} 

Finally, we want to report on an explorative investigation we made. We focused on the impact of edges  intensity on posts virality. The intensity of vertical/horizontal/diagonal
edges 
was computed using Gaussian filters, based on code used in~\cite{Uijlings2010} in the context of real-time
visual concept classification.
The probability density of the average edges intensity follows a gaussian-like  distribution, 
with mean of about 0.08 (both for horizontal and vertical edges). We divided images into two groups: those having an average edge 
intensity below the sample mean, and those having an average edge intensity above the  mean.
Results showed that images with horizontal edge intensity below the sample mean are far more viral on the plusoners and replies indexes, 
while vertical are less discriminative. 
Results for horizontal hedges are as follows: 
$\hat{F}_{plus}(75)$ = 0.36 vs. 0.22, 
$\hat{F}_{repl}(50)$ = 0.27 vs. 0.14, 
$\hat{F}_{resh}(10)$ = 0.25 vs. 0.29, 
K--S test $p<0.001$.
While these results do not have an intuitive explanation, 
they clearly show that there is room for further investigating the impact of edges. 

\subsection{Virality Indexes Correlation} 

From the analyses above, virality indexes seem to ``move together" (in particular plusoners and replies) while resharers appear to 
indicate a different phenomenon. We hypothesize that plusoners and replies can be considered as a form of endorsement, while reshares are 
a form of self-representation. This explains why, for example, pictures
containing faces are endorsed but not used for self-representation by VIPs' followers. On the 
contrary, animated images that  usually contain funny material are
more likely to provoke reshares for followers' self-representation. In fact, people usually tend to represent 
themselves with positive feelings rather than negative ones (especially popular users, see \cite{querciamood}), and positive moods appear to be associated with social interactions \cite{vittengl1998time,de2012not}. 

\begin{table}[h]
\caption{Virality indexes correlation on the datasets}
\begin{center}
\begin{tabular}{l|r|r}
\hline
 & \multicolumn{1}{c|}{Pearson} & \multicolumn{1}{c}{MIC} \\ \hline
\multicolumn{ 3}{c}{\textbf{Static images}} \\ \hline
plusoners vs. replies & 0.723 & 0.433 \\
plusoners vs. resharers & 0.550 & 0.217 \\
replies vs. resharers & 0.220 & 0.126 \\ \hline
\multicolumn{ 3}{c}{\textbf{Animated Images}} \\ \hline
plusoners vs. replies & 0.702 & 0.304 \\ 
plusoners vs. resharers & 0.787 & 0.396 \\ 
replies vs. resharers & 0.554 & 0.205 \\ \hline
\multicolumn{ 3}{c}{\textbf{Text Only}} \\ \hline
plusoners vs. replies & 0.802 & 0.529 \\ 
plusoners vs. resharers & 0.285 & 0.273 \\
replies vs. resharers & 0.172 & 0.185 \\ \hline
\end{tabular}
\label{tab:virality-correlation}
\end{center}
\end{table}

This is supported also by the correlation analysis of the three virality indexes, reported in Table \ref{tab:virality-correlation}, made on the various datasets 
we exploited. In this analysis we used both the Pearson coefficient and the recent Maximal Information Coefficient (MIC), considering  plusoners $\leq$ 1200, replies $\leq$ 400 e
resharers $\leq$ 400. 
MIC is a measure of dependence introduced in \cite{reshef2011detecting} and it is part of the Maximal Information-based 
Nonparametric Exploration (MINE) family of statistics. MIC is able to capture variable relationships of different nature, 
penalizing similar levels of noise in the same way. In this study we use the 
Python package \emph{minepy} \cite{albanese2013minerva}. 

In particular, from Table \ref{tab:virality-correlation} we see that: plusones and replies always 
have a high correlation while replies and resharers always correlate low. Plusoners and reshares, that have a mild correlation 
in most cases, correlate highly when it comes to funny pictures, i.e. animated ones. This can be explained by a specific ``procedural'' effect: 
the follower expresses his/her appreciation for the funny picture and, after that, he/she reshares the content. Since resharing implies also writing 
a comment in the new post, the reply is likely not to be added to the original VIP's post. 

In Table~\ref{dataset-findings} we sum up the main findings of the paper, comparing the various CCDFs: animated images and infographics have much higher
probability of being reshared, while colored images or images containing faces have higher probability of being appreciated or commented. Finally, black and white pictures
(grayscale) turn out to be the least ``viral" on Google+.  

\begin{table}[htdp]
\caption{Summary of main findings of the analysis.}
\begin{center}
\begin{tabular}{l|rrr}
\hline
 &$\hat{F}_{plus}(75)$ & $\hat{F}_{repl}(50)$ & $\hat{F}_{resh}(10)$\\ 
 \hline
very bright & 0.18 & 0.12 & \textbf{0.33}\\ 
grayscale & 0.21 & 0.11 & 0.28\\ 
color & \textbf{0.31} & \textbf{0.24} & 0.27\\ 
animated & 0.17 & 0.08 & \textbf{0.48}\\ 
one-face $>10\%$ area&\textbf{0.35} & \textbf{0.30} & 0.23\\ 
\hline
\end{tabular}
\end{center}
\label{dataset-findings}
\end{table}

\section{User Analysis} 

Finally, we investigate if there is any relevant interaction between images characteristics and VIP's typology. 
In Table~\ref{dataset-usercategory} we report demographic details\footnote{\emph{No Category} denotes users that do not provide any 
personal information and for which it was not possible to trace back their category; \emph{Not Available} denotes seven accounts that were no
more publicly accessible when we gathered demographic info; \emph{Other} denotes very rare and unusual category definitions. 
The \emph{Neutral} gender refers to pages afferent to "non-humans" like products, brands, websites, firms, etc.} on the Google+ dataset, as provided by the users in
their profile pages.  

\begin{table}[htdp]
\caption{User demographics in the Google+ dataset.}
\begin{center}
\begin{tabular}{l|c|c|c|c}
\hline
\textbf{User-category}&\textbf{Female (\%)}&\textbf{Male (\%)}&\textbf{Neutral (\%)}&\textbf{Total (\%)}\\
\hline
Technology & 35 (\emph{19\%}) & 110 (\emph{61\%}) & 36 (\emph{20\%}) & 181 (\emph{19\%})\\
\hline
Photography & 41 (\emph{24\%}) & 130 (\emph{76\%}) & 1 (\emph{1\%}) & 172 (\emph{18\%}) \\
\hline
Music & 96 (\emph{59\%}) & 48 (\emph{29\%}) & 19 (\emph{12\%}) & 163 (\emph{17\%}) \\
\hline
Writing & 26 (\emph{21\%}) & 76 (\emph{63\%}) & 19 (\emph{16\%}) & 121 (\emph{13\%}) \\
\hline
Actor & 21 (\emph{36\%}) & 34 (\emph{59\%}) & 3 (\emph{5\%}) & 58 (\emph{6\%}) \\
\hline
Entrepreneur & 12 (\emph{29\%}) & 29 (\emph{71\%}) & - & 41 (\emph{4\%}) \\
\hline
Sport & - & 22 (\emph{55\%}) & 18 (\emph{45\%}) & 40 (\emph{4\%}) \\
\hline
Artist & 11 (\emph{31\%}) & 21 (\emph{60\%}) & 3 (\emph{9\%}) & 35 (\emph{4\%}) \\
\hline
TV & 8 (\emph{24\%}) & 11 (\emph{33\%}) & 14 (\emph{42\%}) & 33 (\emph{3\%}) \\
\hline
Company & - & - & 28 (\emph{100\%}) & 28 (\emph{3\%}) \\
\hline
Website & - & - & 23 (\emph{100\%}) & 23 (\emph{2\%}) \\
\hline
Politician & - & 19 (\emph{86\%}) & 3 (\emph{14\%}) & 22 (\emph{2\%}) \\
\hline
No Category & 6 (\emph{43\%}) & 8 (\emph{57\%}) & - & 14 (\emph{1\%}) \\
\hline
Organization & - & - & 9 (\emph{100\%}) & 9 (\emph{1\%}) \\
\hline
Not Available & - & - & 7 (\emph{100\%}) & 7 (\emph{1\%}) \\
\hline
Other & 1 (\emph{33\%}) & 2 (\emph{67\%}) & - & 3 (\emph{0\%}) \\
\hline
\hline
\textbf{Total} & \textbf{257 (\emph{27\%})} & \textbf{510 (\emph{54\%})} & \textbf{183 (\emph{19\%})} & \textbf{950 (\emph{100\%})} \\
\hline
\end{tabular}
\end{center}
\label{dataset-usercategory}
\end{table}

In order to investigate 
possible user category effects in our dataset --- that is, if our analyses are also influenced by
the type of user posting images rather than by the actual content solely, we evaluated the entropy for each image category over the 16 user categories (as defined in
Table~\ref{dataset-usercategory}). 
In Table~\ref{dataset-kl} we report the contingency table of image-category entropy distributions over user-categories. 
Looking at the Kullback-Leibler (KL) divergence of
specific image categories with respect to the reference distribution (i.e., taken as the total number of images posted by each 
user-category), we observe very few but interesting
effects due to specific user-categories. 


\begin{table*}[htdp]
\caption{Contingency table of image-category distributions over user-categories.}
\begin{center}
\resizebox{\linewidth}{!}{
\begin{tabular}{l|c|c|c|c|c|c|c|c|c|c|}
\hline
\textbf{User-category}&\textbf{Grayscale}&\textbf{Colored}&\textbf{High Brightness}&\textbf{Low Brightness}&\textbf{Containing Face}&\textbf{Containing No
Face}&\textbf{Squared}&\textbf{Vertical}&\textbf{Horizontal}&\textbf{Total}\\
\hline
No Category & 7\% & 6\% & 9\% & 6\% & 5\% & 7\% & 4\% & 5\% & 7\% & 6\% \\
Actor & 4\% & 6\% & 5\% & 5\% & 8\% & 5\% & 5\% & 6\% & 5\% & 5\% \\
Artist & 5\% & 6\% & 7\% & 6\% & 6\% & 6\% & 5\% & 7\% & 6\% & 6\% \\
Company & 0\% & 1\% & 1\% & 1\% & 1\% & 1\% & 1\% & 1\% & 1\% & 1\% \\
Entrepreneur & 8\% & 7\% & 6\% & 7\% & 7\% & 7\% & 8\% & 5\% & 8\% & 7\% \\
Music & \textbf{3\%} & 16\% & \textbf{3\%} & 16\% & 19\% & 12\% & 15\%
& 29\% & 8\% & 14\% \\
Not Available & 0\% & 0\% & 0\% & 0\% & 0\% & 0\% & 0\% & 0\% & 0\% & 0\% \\
Organization & 0\% & 0\% & 0\% & 0\% & 0\% & 0\% & 0\% & 0\% & 0\% & 0\% \\
Other & 0\% & 0\% & 0\% & 0\% & 0\% & 0\% & 2\% & 0\% & 0\% & 0\% \\
Photography & \textbf{31\%} & 19\% & \textbf{9\% }& 22\% & 15\% & 23\%
& 23\% & 14\% & 23\% & 20\% \\
Politician & 0\% & 0\% & 0\% & 0\% & 0\% & 0\% & 0\% & 0\% & 0\% & 0\% \\
Sport & 0\% & 3\% & 1\% & 3\% & 4\% & 2\% & 2\% & 2\% & 3\% & 2\% \\
Technology & 27\% & 22\% & \textbf{40\%} & 20\% & 19\% & 24\% & 16\% &
18\% & 25\% & 22\% \\
TV & 1\% & 2\% & 1\% & 2\% & 3\% & 1\% & 5\% & 1\% & 2\% & 2\% \\
Website & 1\% & 2\% & 2\% & 2\% & 2\% & 2\% & 1\% & 1\% & 2\% & 2\% \\
Writing & 11\% & 10\% & 17\% & 10\% & 11\% & 10\% & 11\% & 10\% & 11\% & 11\% \\
\hline
\hline
\textbf{KL-divergence} & \textbf{0.173} & 0.002 & \textbf{0.259} & 0.003 &
0.027 & 0.006 & 0.047 & 0.076 & 0.029 \\
\hline
\end{tabular}
}
\end{center}
\label{dataset-kl}
\end{table*}


In particular, while all the KL divergences are very small, two of them (for Grayscale and
 High Brightness, reported in Bold) are an order of magnitude greater than other classes. Interestingly the divergence is explained 
 mainly by the distribution gap in only two User's categories. For High Brightness the gap is mainly given by Technology user 
 category that doubles its probability distribution (from 22\% to 40\%) and Music and Photography that reduce their probability 
 distribution to one third. This divergence from the reference distribution is consistent with the analysis of the content we made in 
 section \ref{Bright-Images}: these images where mainly infographics and screenshots of software programs and social networks 
 (so mainly connected to technology). For Grayscale the gap is mainly given by Photography users category that rises by 50\% its 
 probability distribution and Music, that reduces it to one third. This gap is consistent with the idea, expressed in Section \ref
{Grayscale}, that black-and-white photography is a particular form of art expressivity mainly used by professionals.  

\section{Conclusions} 

We have presented a study, based on a novel dataset of Google+ posts,
showing that perceptual characteristics of an image can strongly
affect the virality of the post embedding it. Considering various kinds
of images (e.g. cartoons, panorama or self-portraits) and related
features (e.g. orientation, animations) we saw that users' reactions
are affected in different ways. We provided a series of analyses 
to explain the underlying phenomena, using three virality
metrics (namely plusoners, replies and resharers). Results suggest
that plusoners and replies ``move together" while reshares indicate a
distinct users' reaction. In particular, funny and informative images have much
higher probability of being reshared but are associated to different image features 
(animation and high-brightness respectively), while colored images or images containing faces 
have higher probability of being appreciated and commented. 

Future work will dig deeper into the
assessment of relations
between visual content and virality indexes, adopting 
multivariate analysis that includes user's categories 
(e.g. which is the viral effect of b/w pictures taken 
by professional photographer as compared to those taken by non professional users).
We will also extend our experimental setup in the
following ways: (a) taking into account compositional features of the
images, i.e. resembling concepts such as the well-known "rule of
thirds"; (b) extracting and exploiting descriptors such as color histograms,
oriented-edges histograms; (c) building upon the vast literature
available in the context of scene/object recognition, dividing our
dataset into specific categories in order to analyse relations between
categories, such as natural images or sport images, and their virality.

\section*{Acknowledgment}

The work of J. Staiano and M. Guerini has been partially supported by the FIRB project S-PATTERNS and the Trento RISE PerTe project, respectively.



%

\bibliographystyle{IEEEtran} 
\bibliography{Persuasive}

\end{document}